\pdfoutput=1
\pdfoutput=1
\documentclass[a4paper,12pt]{article}
\usepackage{jheppub,undertilde}
\usepackage{amsthm}
\usepackage{graphicx}
\usepackage{color}
\usepackage{graphicx,textcomp,float,gensymb,wrapfig, enumitem,comment,dsfont,subfigure,framed,slashed,appendix,wrapfig}

 \newcommand{\be}{\begin{equation}}
\newcommand{\ee}{\end{equation}}
\newcommand{\bea}{\begin{eqnarray}}  
\newcommand{\eea}{\end{eqnarray}}  
\def\code#1{\texttt{#1}}

\preprint{FTUV-16-0707, IFIC/16-46\\}

\title{Sterile Neutrino portal to Dark Matter II: Exact Dark symmetry}

\author[a]{Miguel Escudero,}
\author[a]{Nuria Rius}
\author[b]{and Ver\'onica Sanz}

\affiliation[a]{Departamento de F\'isica Te\'orica and IFIC, Universidad de Valencia-CSIC,
C/ Catedr\'atico Jos\'e Beltr\'an, 2, E-46980 Paterna, Spain}

\affiliation[b]{Department of Physics and Astronomy, University of Sussex, Brighton BN1 9QH, UK\\}

\abstract{We analyze a simple extension of the Standard Model (SM)
with a dark sector composed of a scalar and a fermion, both singlets under the SM 
gauge group but charged under a dark sector symmetry group. 
Sterile neutrinos, which are singlets under both groups, mediate
the interactions between the dark sector and the SM particles,   
and  generate  masses for the active neutrinos via the seesaw mechanism. 
We explore the parameter space region where the 
observed Dark Matter relic abundance is determined 
by the annihilation into sterile neutrinos, both for fermion and scalar  Dark Matter particles.  The scalar Dark Matter case provides an interesting alternative to the usual Higgs portal scenario.
We also study the constraints from direct Dark Matter searches and the 
prospects for indirect detection via sterile neutrino decays to leptons, which may be able to rule out Dark Matter masses below and around 100 GeV.}

\emailAdd{miguel.escudero@ific.uv.es}
\emailAdd{nuria.rius@ific.uv.es}
\emailAdd{v.sanz@sussex.ac.uk}


\begin{document}
\maketitle
\flushbottom

\section{Introduction}

Dark Matter and neutrino masses provide experimental evidence for physics beyond the Standard Model (SM), and finding a scenario where both phenomena are linked is an exciting possibility. Another hint to a connection between these two sectors comes from the standard mechanisms to generate the Dark Matter relic abundance and neutrino masses, as both seem to require new massive degrees of freedom, with a thermal relic and right-handed neutrinos respectively. 

An obvious possibility would be for right-handed neutrinos to constitute the Dark Matter of the Universe~\cite{Dodelson:1993je}. This option is constrained to a specific region at the keV and small mixing with active neutrinos in the minimal see saw model, but  in extended scenarios a larger parameter space is allowed, for instance in the context of a gauged $B-L$ symmetry \cite{Basak:2013cga, Kaneta:2016vkq}.
Upcoming experiments may be able to exclude or establish whether keV neutrinos are the origin of Dark Matter, see e.g.~\cite{Adhikari:2016bei}.

In this paper we take a different approach, focusing on the fact that heavy neutrinos can mediate between Dark Matter and the SM. We propose a simple extension of the Standard Model 
with a new scalar and fermion, singlets under the SM 
gauge group but charged under a dark sector symmetry group. 
Sterile neutrinos, which are singlets under both groups, are able to mediate
the interactions between the dark sector and the SM particles, as well as
 generate  masses for the active neutrinos via the seesaw mechanism. Therefore, the same coupling that generates neutrino masses after electroweak symmetry breaking, determines the Dark Matter phenomenology. Indeed, Dark Matter annihilation to right-handed neutrinos and subsequent decays to SM particles characterize the computation of the relic abundance as well as indirect detection probes,  respectively. 
 
This minimal setup has been studied in \cite{Macias:2015cna,Gonzalez-Macias:2016vxy} for the case of fermion Dark Matter, under the  assumption that the sterile neutrinos are pseudo-Dirac and heavier than the dark sector particles. Our analysis differs from these previous works in that 1) we explore the region of parameter space where sterile neutrinos are lighter than the dark sector, and therefore the Dark Matter can annihilate into sterile neutrinos and 2) we extend the analysis to the scalar Dark Matter case, which was not considered before.

 In a companion paper~\cite{Escudero:2016tzx}, we have explored an alternative scenario with the dark sector charged under $U(1)_{B-L}$, and both papers provide two distinct possibilities for a sterile neutrino portal to Dark Matter.  
 
The paper is organized as follows. After presenting the set-up of our model in Sec.~\ref{sec:darksymm}, we move onto the constraints from Higgs decays and direct Dark Matter searches in Sec.~\ref{Higgs}. We describe the calculation of the annihilation cross section in Sec.~\ref{sec:Relic} where we impose constraints from the relic abundance of Dark Matter. These results are then linked to indirect detection probes via sterile neutrinos decays to leptons in Sec.~\ref{sec:ID}. We conclude in Sec.~\ref{sec:Concls} by summarizing our main findings. 

\section{Exact dark symmetry }~\label{sec:darksymm}
This portal is based upon the assumption that the dark sector contains at least a scalar field 
$\phi$ and a fermion $\Psi$, which are both singlets of the SM gauge group but 
charged under a dark sector symmetry group, $ G_\text{dark}$, 
so that the combination $ \overline{\Psi} \phi$ is a singlet of this hidden symmetry.  
Independently of the nature of the dark group, if
 all SM  particles as well as the sterile neutrinos are  singlets of $ G_\text{dark}$,
    the lighter of the two dark particles turns out to be stable, and therefore it 
    may account for the Dark Matter density of the Universe.  
 \begin{figure}[h!]
\centering
\includegraphics[scale=0.2]{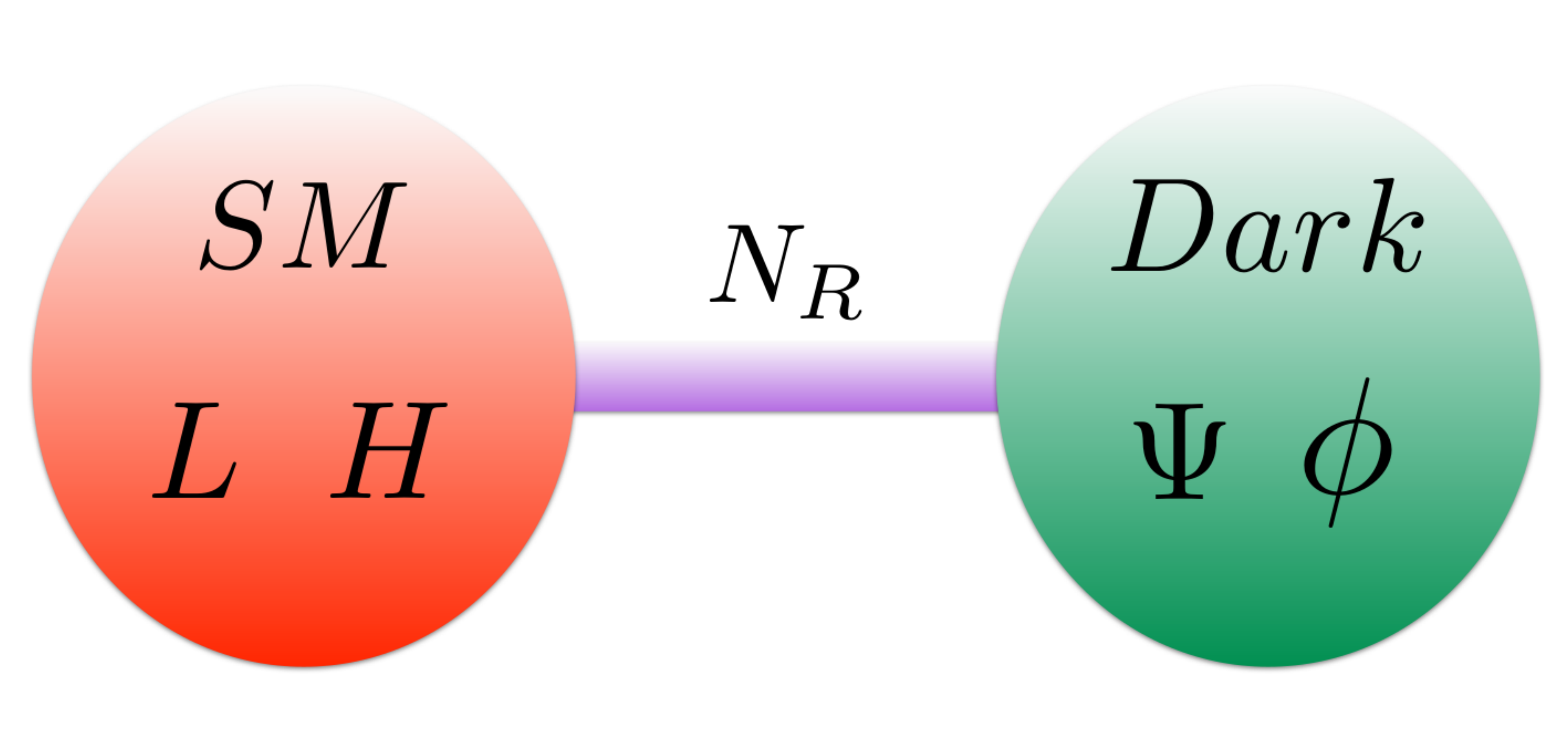}
\end{figure}   
    
 If this were the case then nothing would prevent a term like 
\begin{eqnarray}
\mathcal{L}_\text{int}  	=  -\left( \phi \overline{\Psi}(  \lambda_s + \gamma_5\lambda_p ) N + \phi^\dagger \overline{N} (  \lambda_s - \gamma_5\lambda_p ) \Psi   \right) \label{eq:lag}
\end{eqnarray}
to appear, besides the standard Higgs portal term 
$\lambda_{H\phi}(H^\dagger H) (\phi^\dagger \phi)$  included in the scalar potential,
\bea
\mathcal{L}_\text{scalar} = \mu_H^2 H^\dagger H - \lambda_H (H^\dagger H)^2  -
\mu_\phi^2 \phi^\dagger \phi - \lambda_\phi (\phi^\dagger \phi)^2 - 
\lambda_{H\phi} (H^\dagger H) \, (\phi^\dagger \phi)   ,
\eea
and the term responsible for the generation of neutrino masses
 \bea
\mathcal{L}_{\nu N} = -(Y_{\alpha a} \overline {L}_L^\alpha H N_{Ra} + h.c.) , 
\label{eq:nuN}
\eea
where $\alpha=e,\mu, \tau$ denotes lepton flavour and $a=1\ldots n$, being $n$ the number of 
sterile neutrinos.

For simplicity we do not consider the possibility that the scalar $\phi$ gets a vev, and we restrict the discussion to   the minimal matter content, although there could be more than one set of dark fermions and scalars.

Another simplifying assumption made in this paper is that  the dark symmetry $G_\text{dark}$ is a global symmetry at low energies. We are therefore neglecting the possible phenomenology of $G_\text{dark}$ vector mediators,
e.g., 
if the dark symmetry were local there could also be kinetic mixing among the dark gauge bosons and the SM ones, leading to further SM-dark particle interactions \cite{Heeck:2012bz, deGouvea:2015pea}.
The following discussion will apply as well to this scenario, provided the kinetic mixing is negligible. Nevertheless, the UV structure and stability of Dark Matter depends on whether $G_\text{dark}$ is a true global symmetry or a gauge symmetry. Global symmetries are sensitive to higher-dimensional operators mediated by quantum gravity effects~\cite{Kallosh:1995hi}, e.g. $c_{\Psi} \bar \Psi \tilde H^\dagger \gamma^\mu D_\mu L/M_{Pl}$ or $c_\Phi  \Phi F_{\mu\nu} F^{\mu\nu}/M_{Pl}$, and could lead to disastrous decay of Dark Matter unless $c_{\Psi,\Phi}\ll 1$~\cite{Cata:2014sta,Mambrini:2015sia,Cata:2016dsg}.

Regarding the neutrino sector, light neutrino masses are generated via TeV scale type I seesaw mechanism, 
that we briefly review in the following.
We denote $\nu_\alpha$ the active neutrinos and $N_s$ the sterile ones. 
After electroweak symmetry breaking, the neutrino mass matrix in the basis 
$(\nu_\alpha, N_s)$ is given by
\be
\label{eq:numass}
{\cal M}_\nu = \left(
\begin{array}{cc}
0 & m_D \\
m_D^T & m_N
\end{array}
\right) \ , 
\ee
where $m_D = Y v_H/\sqrt{2}$ and $Y_{\alpha s}$ are the Yukawa couplings. 

The matrix 
${\cal M}_\nu$ can be diagonalized by a unitary matrix $U$, so that 
\be
{\cal M}_\nu = U^* \, Diag(m_\nu,M) \, U^\dagger \ , 
\ee
where 
$m_\nu$ is the diagonal matrix with the three lightest eigenvalues of ${\cal M}_\nu$,  
of order $m_D^2/m_N$, 
and $M$ 
contains the heavier ones, of order $m_N$. 
 
The mass eigenstates ${\bf n}=(\nu_i,N_h$) are related to the active and sterile neutrinos, 
($\nu_\alpha$, $N_{s}$), by
\bea
\left(\begin{array}{c}\nu_\alpha \\ N_s \end{array}\right)_L  =  U^* \,  
\left(\begin{array}{c}\nu_i  \\ N_h \end{array}\right)_L  \ .
\eea
The unitary matrix $U$ can be written as 
\be 
\label{eq:mixing}
U = \left(
\begin{array}{cc}
U_{\alpha i } & U_{\alpha h}  \\
U_{s i } & U_{s h } 
\end{array}
\right) \ , 
\ee
where, at leading order in the seesaw expansion parameter, ${\cal O}(m_D/m_N)$:

\bea
U_{\alpha i } &=& [U_\text{PMNS} ]_{\alpha i} \qquad   U_{sh} = I 
\nonumber \\
U_{\alpha h } &=&  [m_D m_N^{-1}]^*_{\alpha h}
\\
U_{s i} &= & - [m_N^{-1} m_D^T \, U_\text{PMNS}]_{si} \ .
\nonumber 
\eea
Notice that at this order the states $N_h$ and $N_s$ coincide, so we identify them in the 
rest of this paper. 

Neglecting the mixing between the CP-even scalars, the Yukawa coupling of the SM-like  Higgs field  $h$  to the neutrinos can be written as 
\cite{Pilaftsis:1991ug}:
\be
{\cal L}_Y = - \frac{ h}{2 v_H}  \bar {\bf n}_i [ ( m_i + m_j) Re (C_{ij} )+ 
i \gamma_5 (m_j - m_i) Im (C_{ij}) ]  {\bf n}_j \ ,
\ee
where the indices $i,j$ refer to the light neutrinos $\nu_i$ 
for $i,j =1,2,3$ and to $N_h$ for $i,j =4,5,6$, and the matrix $C$ can be written in terms of
the mixing matrix $U$ as:
\be
\label{eq:cij}
C_{ij} = \sum_{\alpha=1}^{3} U_{\alpha i} U^*_{\alpha j} \ .
\ee

A variation of this scenario has been analyzed in \cite{Macias:2015cna,Gonzalez-Macias:2016vxy}, where the sterile neutrinos are assumed to be pseudo-Dirac and \emph {heavier} than the dark sector particles, $\phi, \Psi$.
Thus, they can be integrated out and generate at tree level the effective dimension five operator 

\be
{\cal O}^{(5)} = (\overline{\Psi} \phi) (\widetilde H^\dagger \ell) \ , 
\ee 
which after the SM Higgs doublet acquires a vev leads to the interaction 
${\cal O}^{(5)} = (\overline{\Psi} \phi) \nu_L v_H/\sqrt{2}$, involving a SM left-handed 
neutrino. This  limit  is in fact a (light) neutrino portal to Dark Matter.
Assuming that the fermion $\Psi$ is the Dark Matter,  
the model can accommodate current experimental and  
observational constraints if $M_\psi$ is below $\sim$ 35 GeV, or it is in a resonant region of 
the Higgs or $Z$ boson, or the dark scalar and dark fermion are almost degenerate.

Our analysis is  complementary, since we focus on a different region of the model parameter space: we assume that the sterile neutrinos are \emph{lighter} than  the Dark Matter
and therefore the annihilation channel to $N N $ is open.
Furthermore, we study both fermion and scalar Dark Matter. 
Although the scalar Dark Matter case falls among the  class of Higgs portal models that have been extensively studied~\cite{Patt:2006fw,Kim:2006af,MarchRussell:2008yu,Kim:2008pp,Ahlers:2008qc,Feng:2008mu,Andreas:2008xy,Barger:2008jx,Kadastik:2009ca,
Kanemura:2010sh,Piazza:2010ye,Arina:2010an,Low:2011kp,Djouadi:2011aa,Englert:2011yb,Kamenik:2012hn,Gonderinger:2012rd,Lebedev:2012zw,Craig:2014lda,Queiroz:2014yna}, it is worth to explore whether the new annihilation channel into $N N $ allows to 
obtain the observed relic density in regions that are excluded in the standard Higgs 
portal framework.

In the following sections we describe the current constraints  on the above scenario and the results of our 
numerical analysis, based on a Monte Carlo scan over the free parameters 
($m_\Psi, m_\phi,  m_N, \lambda_s, \lambda_{H\phi}$)  in logarithmic scale, 
restricting the values of the couplings and masses to the ranges displayed in Table~\ref{tab:RUN}.
We present the analytic results for arbitrary Dark Matter - sterile neutrino couplings $\lambda_s,\lambda_p$, 
but  for the numerical implementation we have chosen $\lambda_p=0$, since as explained in 
sec.~\ref{sec:Relic}, in this case strong constraints can be set from indirect Dark Matter searches. 

We made use of \code{LanHep}~\cite{Semenov:2008jy} and \code{micrOMEGAs}~\cite{Belanger:2013oya} in order to obtain the correct relic abundance, Higgs decays and today's annihilation cross section. We calculate $10^6$ points that match the {\it Planck} constraint on the Dark Matter abundance at $3\sigma$~\cite{Ade:2015xua}, namely $\Omega h^2 = 0.1198 \pm 0.0045$. 

\begin{table}[t]
\begin{center}
\begin{tabular}{cccccc}
\hline\hline
 Case & $m_{\Psi}$ (GeV)  & $m_\phi$ (GeV) &$m_N$ (GeV) & $\lambda_s $ & $ \lambda_{H\phi} $\\
\hline
Fermion DM & $1-2\times10^{3}$ & $1-10^{4} $ & $1-2\times10^{3}$ & $10^{-4}-4\pi$ & $10^{-4}-4\pi$ \\
Scalar DM & $ 1-10^{4} $ & $1-2\times10^{3}$  & $1-2\times10^{3} $ & $10^{-4}-4\pi$ & $10^{-4}-4\pi$ \\
\hline\hline
\end{tabular}
\caption{Explored parameter space in the models.} 
\label{tab:RUN}
\end{center}
\end{table}

\section{Constraints from Higgs decays and direct Dark Matter searches }\label{Higgs}

The enlarged fermion and scalar sectors lead to new decays of the Higgs boson, $h$, 
which can be constrained using the 
ATLAS and CMS limits on the invisible Higgs decay branching fraction: 
\be
\label{eq:h_inv}
{\rm BR_{inv}} = \frac{\Gamma_{\rm inv}}{\Gamma_{\rm inv} + \Gamma_{\rm SM}} 
< 0.23 \qquad (95 \% {\rm CL})\ ,
\ee 
being the SM Higgs width $\Gamma_{\rm SM} \approx 4$ MeV.

\begin{figure}[t!]
\begin{center}
\includegraphics[width=0.8\textwidth]{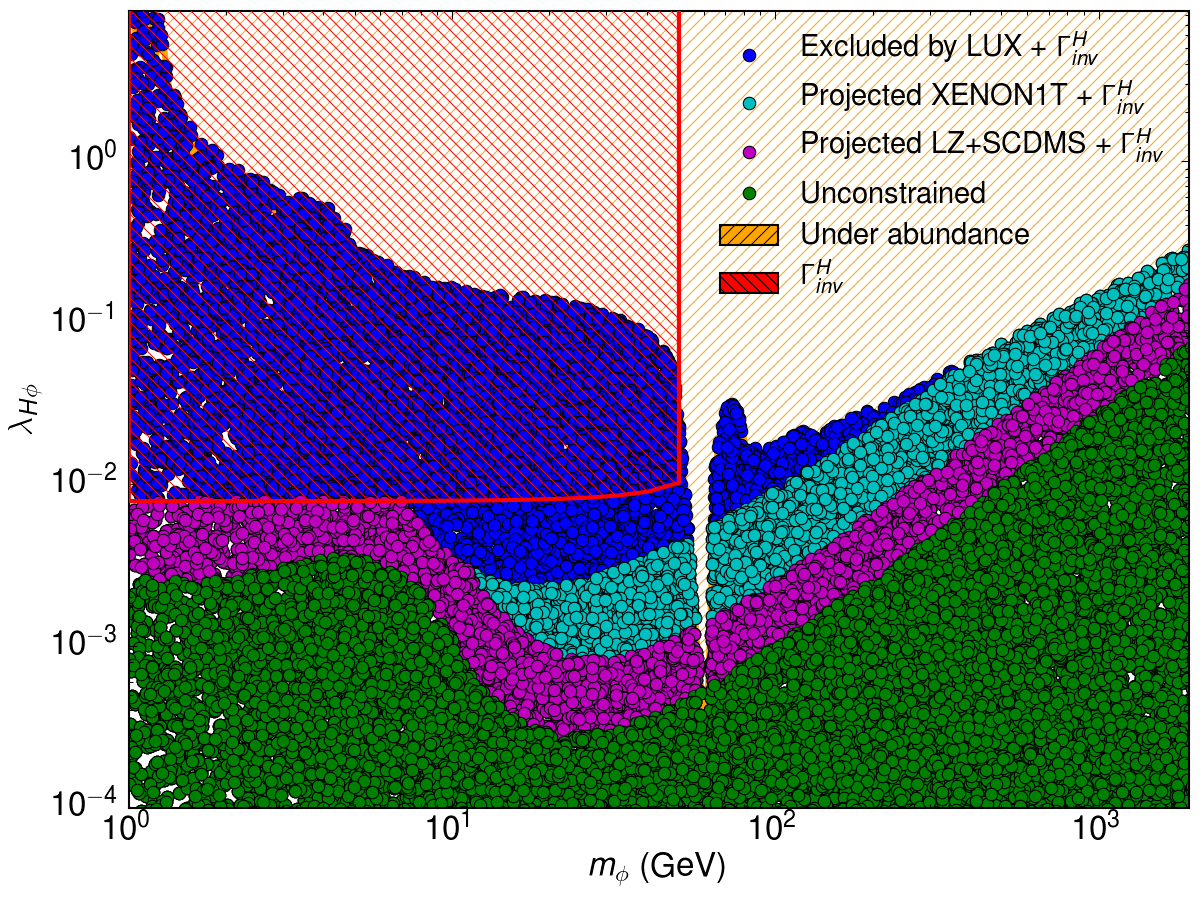} 
 \caption{Constraints on the Higgs portal coupling for scalar DM.}
\label{fig:DirectDetectionS}
\end{center}
\end{figure}

At tree level, there are  two new Higgs decay channels: 
when $m_\phi < m_h/2$, 
the standard decay of the Higgs portal scenarios, $h \rightarrow \phi \phi$ is 
kinematically allowed, contributing to the  
invisible Higgs decay width by:
\be  
\Gamma(h \rightarrow \phi \phi) = \frac{\lambda_{H\phi}^2  v_H^2 }{8 \pi m_h}
\sqrt{1 - \frac{4 m_\phi^2}{m_h^2}}
\ee
We show in Fig. \ref{fig:DirectDetectionS}  the upper limit on the Higgs portal coupling 
$\lambda_{H\phi}$ 
derived from the experimental limit on the
invisible Higgs decay width in Eq.~(\ref{eq:h_inv}), 
as a function of the singlet scalar mass, $m_\phi$.

Moreover,  the Yukawa interaction term $Y \overline {L} H P_R N $ also leads to novel Higgs decay channels into neutrinos. 
The corresponding decay width reads \cite{Pilaftsis:1991ug}:
\be
\label{eq:h_ninj} 
\Gamma(h \rightarrow  {\rm n}_i {\rm n}_j ) = \frac{\omega}{8 \pi m_h} 
\lambda^{1/2}(m_h^2,m_i^2,m_j^2) \left[
S \left(1 - \frac{(m_i+m_j)^2}{m_h^2} \right) 
+ P \left(1 - \frac{(m_i-m_j)^2}{m_h^2} \right) \right ] \ , 
\ee
where $\lambda(a,b,c)$ is the standard kinematic function, $w=1/n!$ for $n$ identical 
final particles and  
the scalar and pseudoscalar couplings are: 
\be 
S = \frac{1}{v_H^2} [ (m_i+m_j) Re(C_{ij})]^2  
\qquad , \qquad
P=  \frac{1}{v_H^2} [ (m_j- m_i) Im (C_{ij})]^2   \ , 
\ee
with $C_{ij}$  defined in Eq.~(\ref{eq:cij}).

The largest branching ratio is for the decay into one light and one heavy neutrino
\cite{Gago:2015vma}:
\be 
\Gamma (h \rightarrow \nu N) = \frac{m_N^2}{8 \pi v_H^2}   
\left( 1 - \frac{m_N^2}{m_h^2} \right)^2 m_h | C_{\nu N }|^2  \ .
\ee
The attainable values for the above branching fractions have been analyzed in 
\cite{Gago:2015vma}, for the case of two heavy neutrinos, 
parameterizing the Yukawa couplings in terms of the observed light neutrino 
 masses and mixing angles, and a complex orthogonal matrix.
 After imposing the relevant constraints from neutrinoless double beta decay, lepton flavour violating processes and direct searches of heavy neutrinos, they find that branching 
 ratios of $h \rightarrow \nu_i N_a$ larger than 
  $10^{-2}$ are generally ruled out for heavy neutrino masses 
 $M_N \leq 100$ GeV, and typically they are much smaller, due to the  tiny Yukawa couplings
 required to fit  light neutrino masses with sterile neutrinos at the electroweak scale.
 Therefore, the contribution of such decay modes to the Higgs decay width is 
 negligible, and they do not alter the bounds discussed above.

\begin{figure}[t!]
\centering
 \begin{tabular}{cc}
\includegraphics[width=0.45\textwidth]{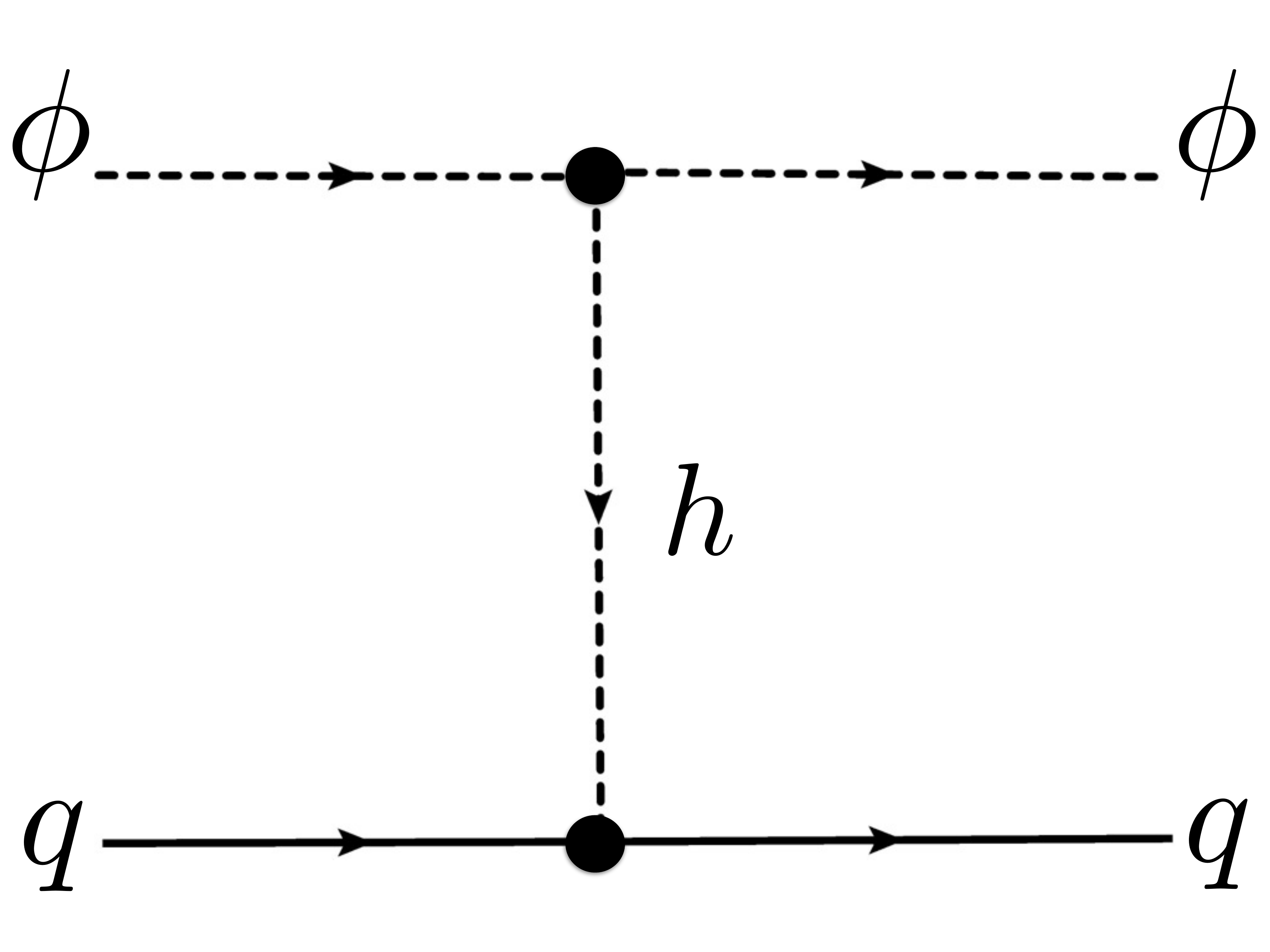} & \includegraphics[width=0.49\textwidth]{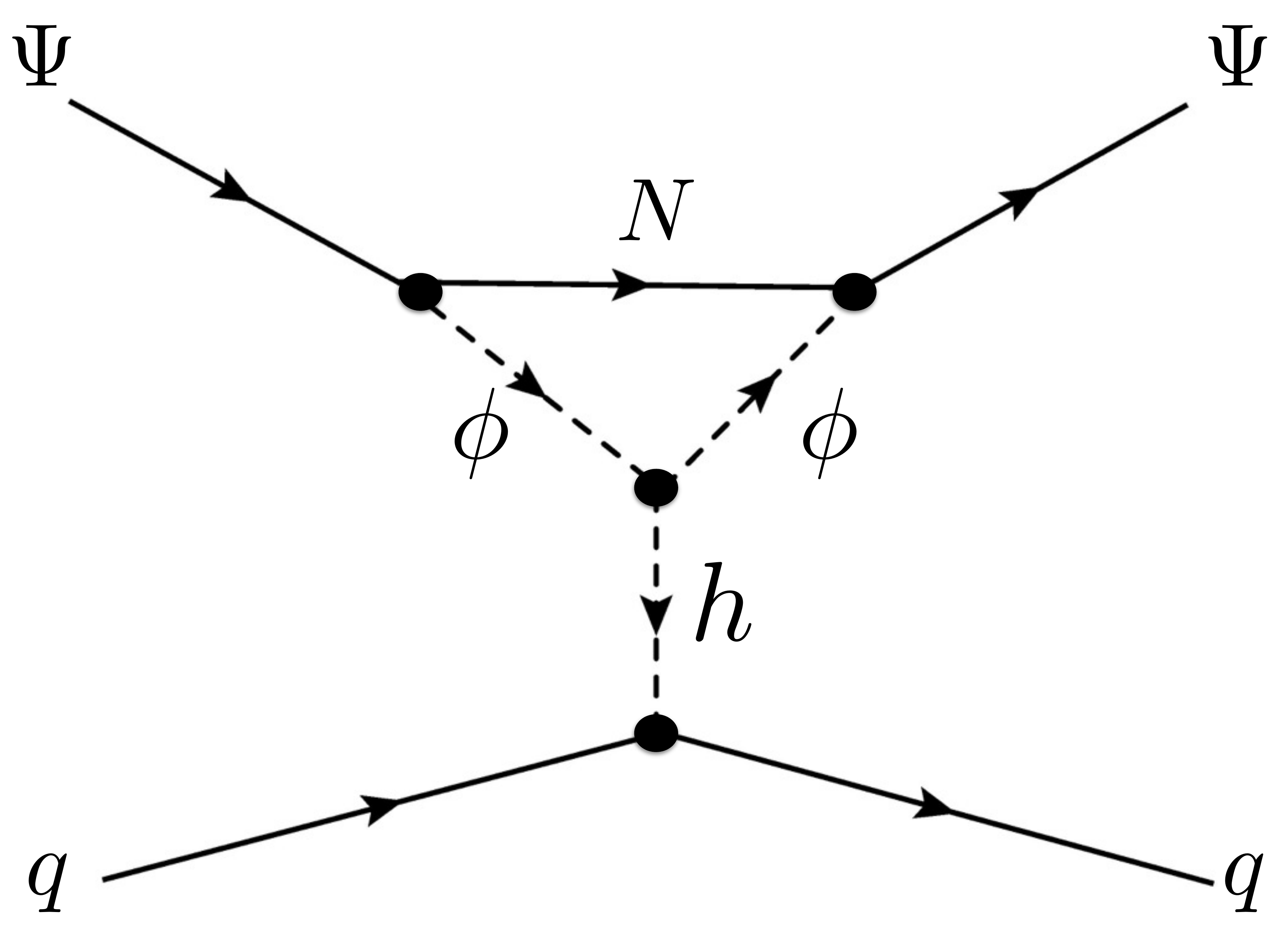} \\
\end{tabular}
 \caption{ {\it Left, Right:} elastic cross section diagrams for the scalar and fermion dark matter cases, respectively.} 
\label{fig:1loopDD}
\end{figure}

At one loop,  the $d=5$ Higgs portal operator   
$\overline{\Psi} \Psi (H^\dagger H)$ is generated (unless the coupling of the Dark Matter to the 
dark scalar and sterile neutrinos is chiral, i.e., $\lambda_s = \lambda_p$ in Eq. ~(\ref{eq:lag})) 
with a coefficient given by \cite{Gonzalez-Macias:2016vxy}:
\begin{equation}
\label{eq:1Loopcontri}
\lambda_{H \Psi}^{eff}=  \lambda_{H \phi} \frac{(\lambda_s^2 - \lambda_p^2)}{16\pi^2} 
 \frac{m_N}{(m_\phi^2-m_N^2)^2}
 \left( m_\phi^2-m_N^2+m_N^2 \log{\frac{m_N^2}{m_\phi^2}} \right)  \ . 
\end{equation}
Thus, when $m_\Psi < m_h/2$ the invisible decay $h \to \Psi \Psi$ is also allowed with partial decay width 
\be
\Gamma(h \rightarrow \Psi \Psi) = \frac{(\lambda_{H  \Psi}^{eff})^2}{8 \pi}
\left( 1 - \frac{4 m_\Psi^2}{m_h^2} \right)^{3/2} m_h  \ ,
\ee
and  the current limit on the invisible Higgs decay branching ratio only leads to an ${\cal O}(1)$ constrain on $\lambda_{H \phi}$, depending  on the values of the remaining free parameters, namely
$\lambda_s, \lambda_p, m_N$ and $m_\phi$. Notice however that if 
$m_\phi < m_h/2$, the strong constraints from the invisible Higgs decay 
$h \to \phi \phi$ shown in Fig. \ref{fig:DirectDetectionS} will apply as well.

\begin{figure}[t!] 
\includegraphics[width=0.98\textwidth]{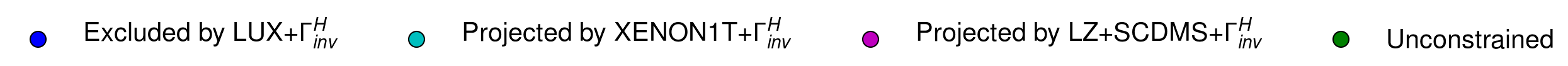} 
\begin{tabular}{cc}
\includegraphics[width=0.49\textwidth]{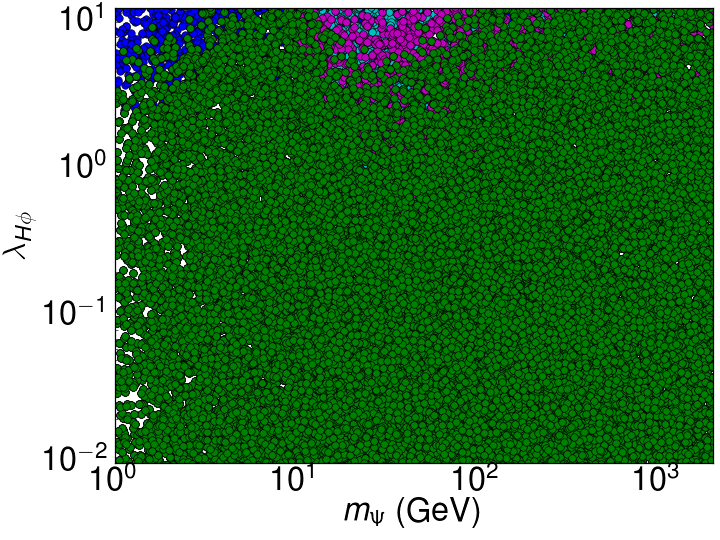} & \includegraphics[width=0.49\textwidth]{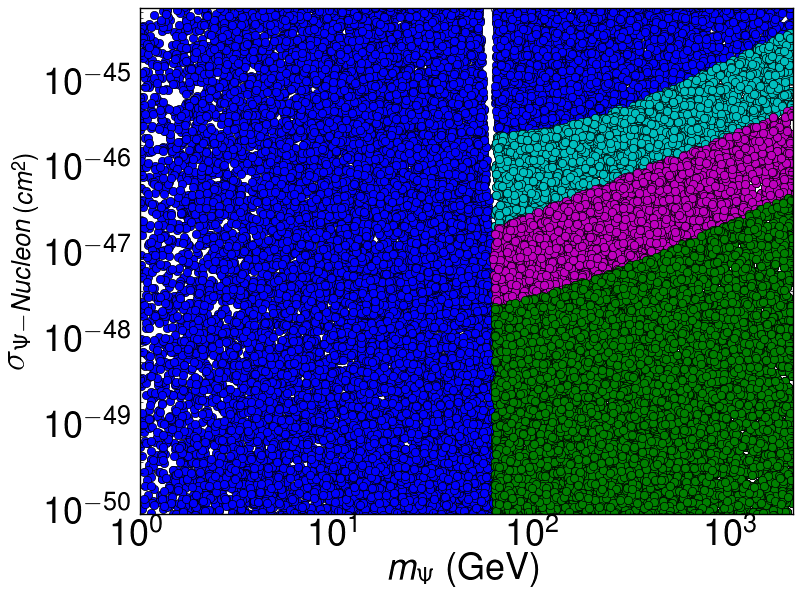} \\
\end{tabular}
 \caption{Constraints on the Higgs portal coupling for fermion DM.}
\label{fig:DirectDetectionF}
\end{figure}

Concerning the bounds from direct DM searches, they also depend on 
which of the dark particles is lighter, and therefore stable.
In order to implement such bounds
we shall assume that the DM relic density is as determined by CMB measurements, since this requirement is always fulfilled  in our scenario for both scalar and fermion DM, as we will see in the next section.

If DM is the dark fermion, $\Psi$, it only interacts with the SM quarks at one loop level 
(see Fig.~\ref{fig:1loopDD}), via the induced Higgs portal operator 
$\overline{\Psi} \Psi (H^\dagger H)$ just discussed, and 
therefore the bounds from direct detection are quite weak. However, since the interaction to quarks is mediated through the Higgs, the scattering will always be spin independent. We refer the reader for the actual matrix elements to~\cite{Kumar:2013iva}. In Fig.~\ref{fig:DirectDetectionF} we show the excluded region by the invisible Higgs decay and current LUX~\cite{Akerib:2012ys,LUX2016:ref} results (dark blue points), as well as the expected excluded region by XENON1T~\cite{Aprile:2015uzo} (light blue) and LZ~\cite{Akerib:2015cja,McKinsey:2016xhn}+SuperCDMS~\cite{Agnese:2015ywx} (purple). Similar constraints can be set with the current results from the PANDAX experiment~\cite{Tan:2016zwf}.

However, if DM is the dark scalar $\phi$, it interacts with the SM quarks at tree level 
via the Higgs portal coupling, $\lambda_{H\phi}$, and the null results from direct 
searches set strong limits on this parameter. This is illustrated in 
Fig.~\ref{fig:DirectDetectionS}, where we show the allowed values 
 of the Higgs portal coupling $\lambda_{H\phi}$ as a function of the DM mass, $m_\phi$, derived 
from the invisible Higgs decay width plus LUX bounds, as well as the prospects from XENON1T and LZ+SuperCDMS. 
The dark blue points in the usual Higgs portal scenario would be ruled out, except for the upper limit,
since the $\lambda_{H\phi}$ being too small leads to a DM relic density larger than the one determined 
by CMB measurements. In our scenario the alternative annihilation channel into $ N N $ 
provides the correct relic density, but the current constraints from LUX and Higgs invisible decay width
excludes them. We notice that for $m_\phi \gtrsim 300 \, {\rm GeV}$ the usual Higgs portal model still provides the correct relic abundance. However, we find that XENON1T can be sensitive to such scenario for $m_\phi < 2 \, {\rm TeV}$.

\section{Dark Matter relic abundance}\label{sec:Relic}
\begin{figure}[t!]
\centering
 \begin{tabular}{ccc}
\includegraphics[width=0.31\textwidth]{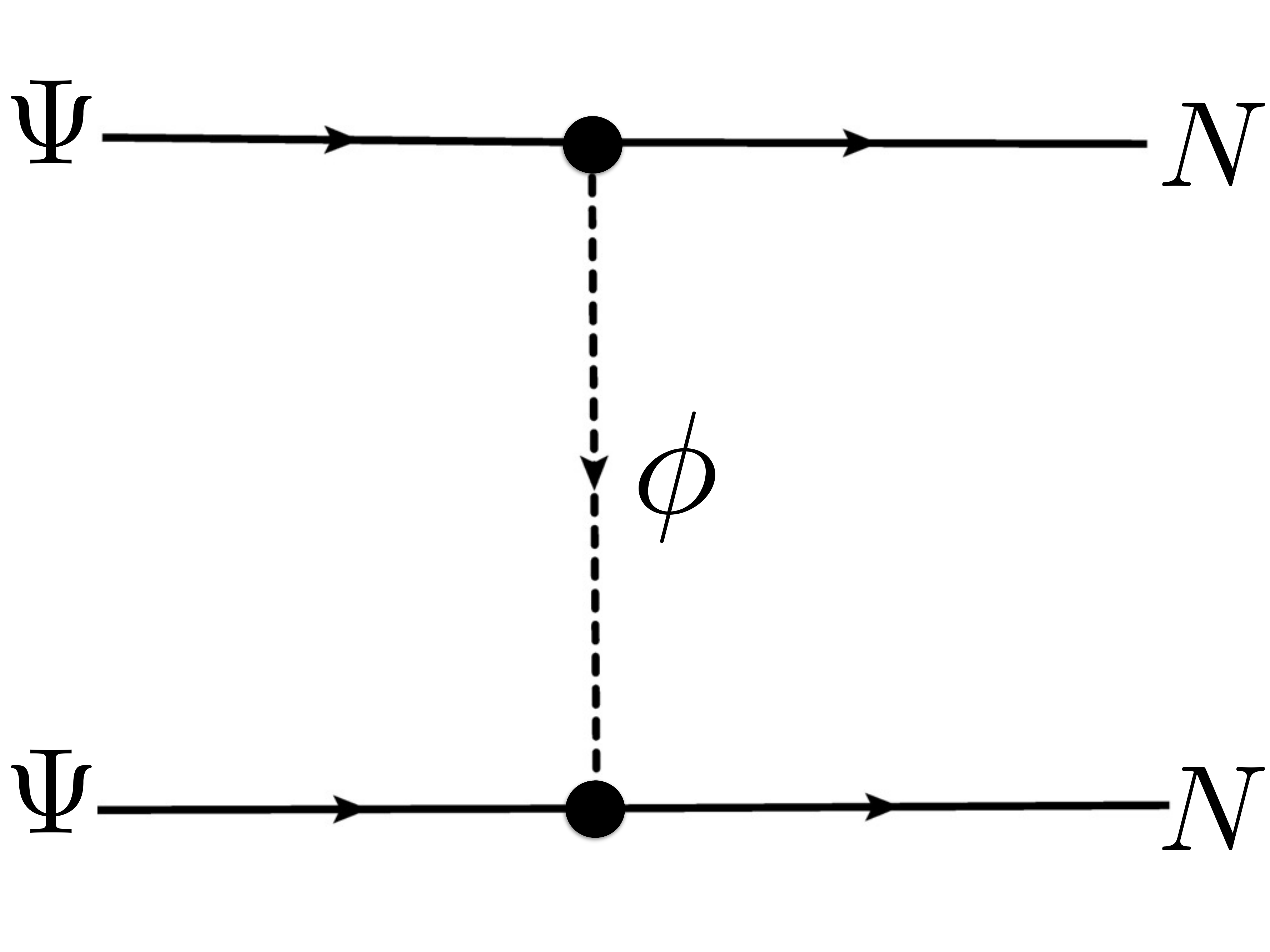} & \includegraphics[width=0.31\textwidth]{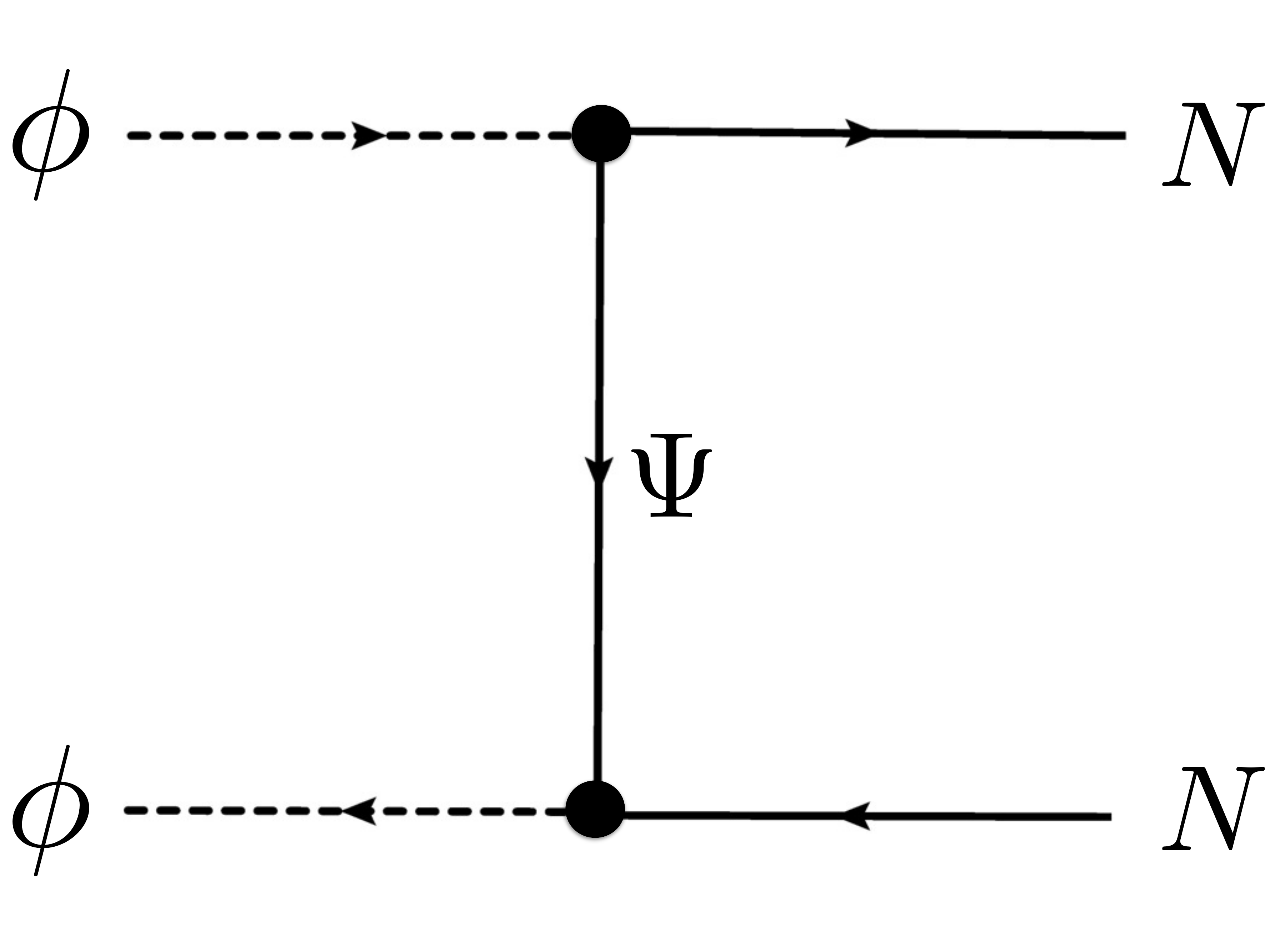} & \includegraphics[width=0.31\textwidth]{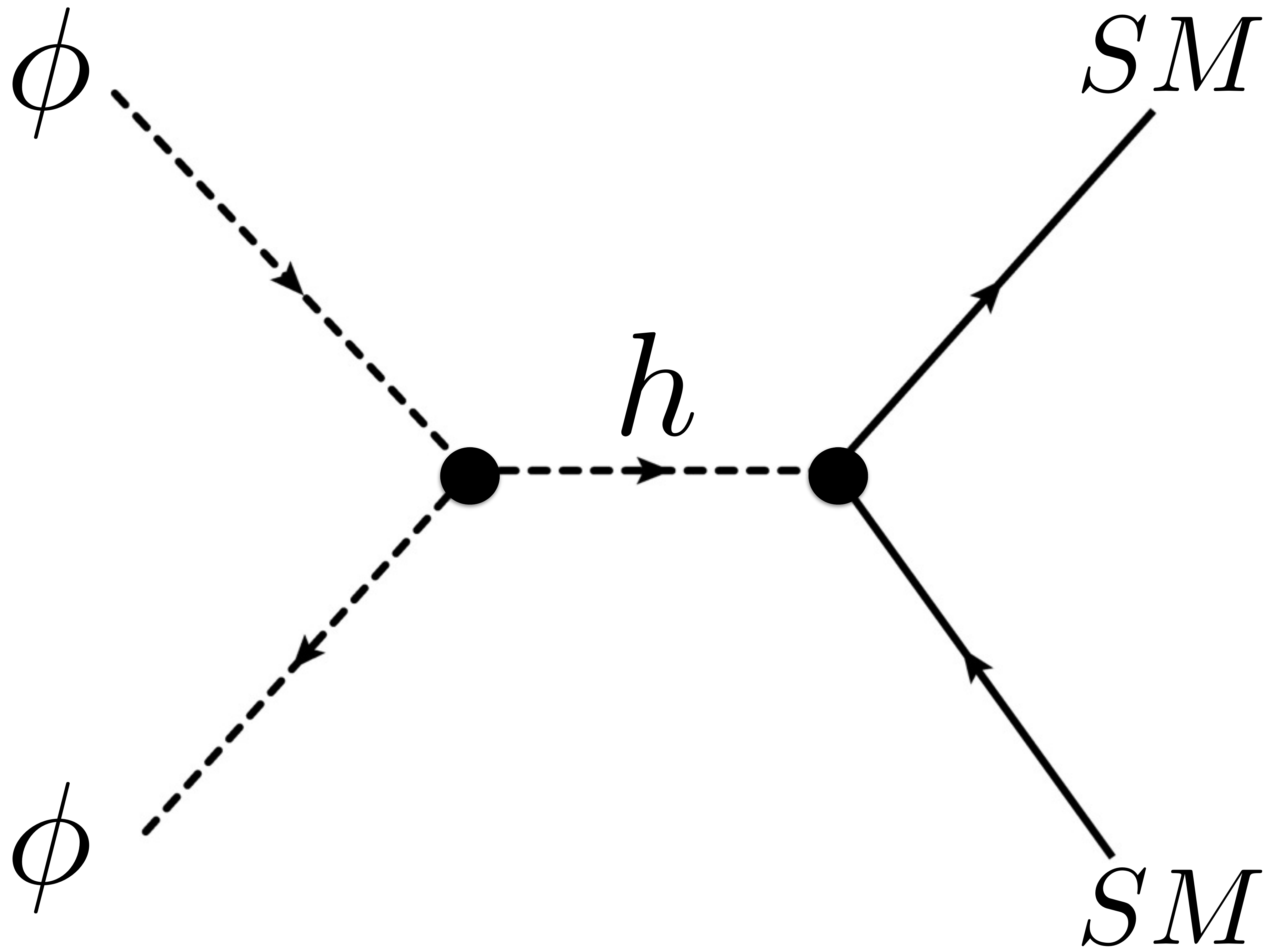}  \\
\end{tabular}
 \caption{Relevant annihilation channels.} 
\label{fig:Anndiagrams}
\end{figure}

\subsection{Thermal history}
In order to discuss the thermal production of Dark Matter in the early Universe we will first describe the thermal history for both the scalar and fermion dark matter scenarios. 
\begin{enumerate}
\item {\bf Fermion Dark Matter $\Psi$: } At very early times $\phi$, $\Psi$ and $N$ are in thermal equilibrium with the standard model via the Higgs portal coupling. The heavy dark particle companion will decay at $T \lesssim m_\phi$ and the dark sector may still be coupled to the standard model bath if the Yukawa couplings of the sterile neutrinos are large enough. If they are small, then the $\Psi$ and $N$ bath will decouple and remain in thermal equilibrium but with a different temperature\footnote{Yet, the actual value of the Yukawa couplings are not known. The naive seesaw expectation is $Y \sim \sqrt{m_\nu m_N}/v_H \sim  4 \times 10^{-8} \sqrt{m_N/(1 \, {\rm GeV})}$ for $m_\nu \sim 0.1$ eV, but larger couplings are consistent with neutrino masses, for instance in the context of inverse seesaw scenarios.}. Then when the temperature of such a bath is $T_D\sim m_\Psi/20$ the dark matter will be produced and the sterile neutrinos will decay at $T_D \lesssim m_N$. In order to check whether the decoupling of the dark sector will modify the production rate it is worth revisiting the production mechanism, see~\cite{Berlin:2016gtr} for a recent discussion on decoupled dark sectors. Since the 
the entropy is separately conserved in both the visible and the dark sectors, 
the standard relic abundance solution is modified approximately by a factor $\sqrt{g_\star^\text{eff}/g_\star}$ where $g_\star$ measures the total number of relativistic degrees of freedom in the SM bath and $g_\star^\text{eff} = g_\star + g_D (T_D/T)^4$ represents the effective number of relativistic species. 
Given that the number of degrees of freedom in the dark sector, $g_D$ is much smaller than $g_\star$ and $T_D/T \sim1$ then $\sqrt{g_\star^\text{eff}/g_\star}$ must be close to one. Thus, a sizeable change in the couplings compared to the case in which both sectors remain in thermal equilibrium is not expected since furthermore $\Omega_{\chi}h^2 \propto 1/\lambda_{s,p}^{4}$. The only caveat to this argument occurs when $m_\chi \gtrsim m_N$, because in that case the sterile neutrinos may have a larger number density than the equilibrium one and in order to generate the same amount of dark matter higher couplings between $\chi$ and $N$ will be needed. This scenario has been recently studied by~\cite{Tang:2016sib} for the precise model proposed in this work. They found that in such region one will need couplings a factor between $1-4$ higher depending on the Yukawa of the sterile neutrinos. Since this change is mild, for our computations we will assume that all species are equilibrium with the standard model.

\item {\bf Scalar Dark Matter $\phi$: } At very early times $\phi$, $\Psi$ and $N$ are in thermal equilibrium with the standard model via the Higgs portal coupling. The heavy dark particle companion will decay at $T \lesssim m_\Psi$ and the dark sector will decouple from the standard model when the dark matter freezes out at $T \sim m_\Psi/20$ and the sterile neutrinos will decouple and decay at $T \lesssim m_N$.    
\end{enumerate}
\subsection{Relic abundance}
In our scenario, the annihilation cross section into two sterile neutrinos 
depends on the nature of the DM particle (scalar, Dirac or Majorana fermion) and the type of coupling (scalar, pseudoscalar). The relevant Feynman diagrams are shown in Fig.~\ref{fig:Anndiagrams}.
For example, let us assume right-handed neutrinos are Majorana, and consider the two options of fermion and scalar Dark Matter: 

\begin{enumerate}
\item {\bf Fermion Dark Matter $\Psi$: } The cross section for fermionic Majorana Dark Matter and complex mediator $\phi$ reads as
\bea
\sigma v_{\Psi \Psi \to NN} =\frac{ (\alpha + \beta \, r_{N \Psi})^2}{4 \pi m_\Psi^2 }   \, \frac{ \sqrt{1-r_{N \Psi}^2}}{(1+r_\phi^2-r_{N \Psi}^2)^2} + {\cal O}(v^2) \label{eq:rel1}
\eea
where $\alpha= \lambda_s^2-\lambda _p^2 $ and $\beta=\lambda_s^2+\lambda _p^2$, $r_{\phi} = m_\phi/m_{\Psi}$, and $r_{N \Psi} = m_N/m_\Psi$. 

One can obtain the case of a Dirac DM particle by in Eq.~(\ref{eq:rel1}) perform the exchange $\alpha \leftrightarrow \beta$. Similarly, the case of a real scalar can be obtained by setting $\lambda_p=0$ in Eq.~(\ref{eq:rel1}), which leads to $\alpha=\beta=\lambda_s^2$ in this expression. 

\item{\bf Scalar Dark Matter $\phi$: } In the case of a real scalar Dark Matter and Dirac mediator $\Psi$ the cross section is as follows,
\bea
 \sigma v_{\phi \phi \to NN}  = \frac{ (\alpha + \beta \, r_{N \Psi})^2}{2 \pi m_\phi^2 }   \, \frac{(1-r_{N \phi}^2)^{3/2}}{(1+r_\Psi^2-r_{N \phi}^2)^2} + {\cal O}(v^2) \ , \label{eq:rel2}
\eea
where $r_\Psi=m_{\Psi}/m_\phi$ and $r_{N \phi} = m_N/m_\phi$.

To obtain the expression for a complex scalar, one can multiply this equation by a factor $1/4$. Similarly, to consider a Majorana mediator one would multiply the expression by a factor 4 and set $\lambda_p$ to zero, $\alpha=\beta$~\footnote{Note that our results agree with the expressions obtained in Ref.~\cite{Berlin:2014tja}, where both fermions were set to be Dirac particles.}.
\end{enumerate}

An important observation is that there are situations where the annihilation cross section at leading order in the relative Dark Matter velocity, $v$, is proportional to the right-handed neutrino mass. For example, the case of a Majorana Dark Matter  with chiral couplings,  $|\lambda_s|=|\lambda_p|$ ($\alpha=0$). In this case when $m_N \ll m_\phi, m_\Psi$  the cross section is effectively
 p-wave, which reduces the sensitivity of indirect detection probes to these scenarios.  
 
 \begin{figure}[t!]
\includegraphics[width=0.98\textwidth]{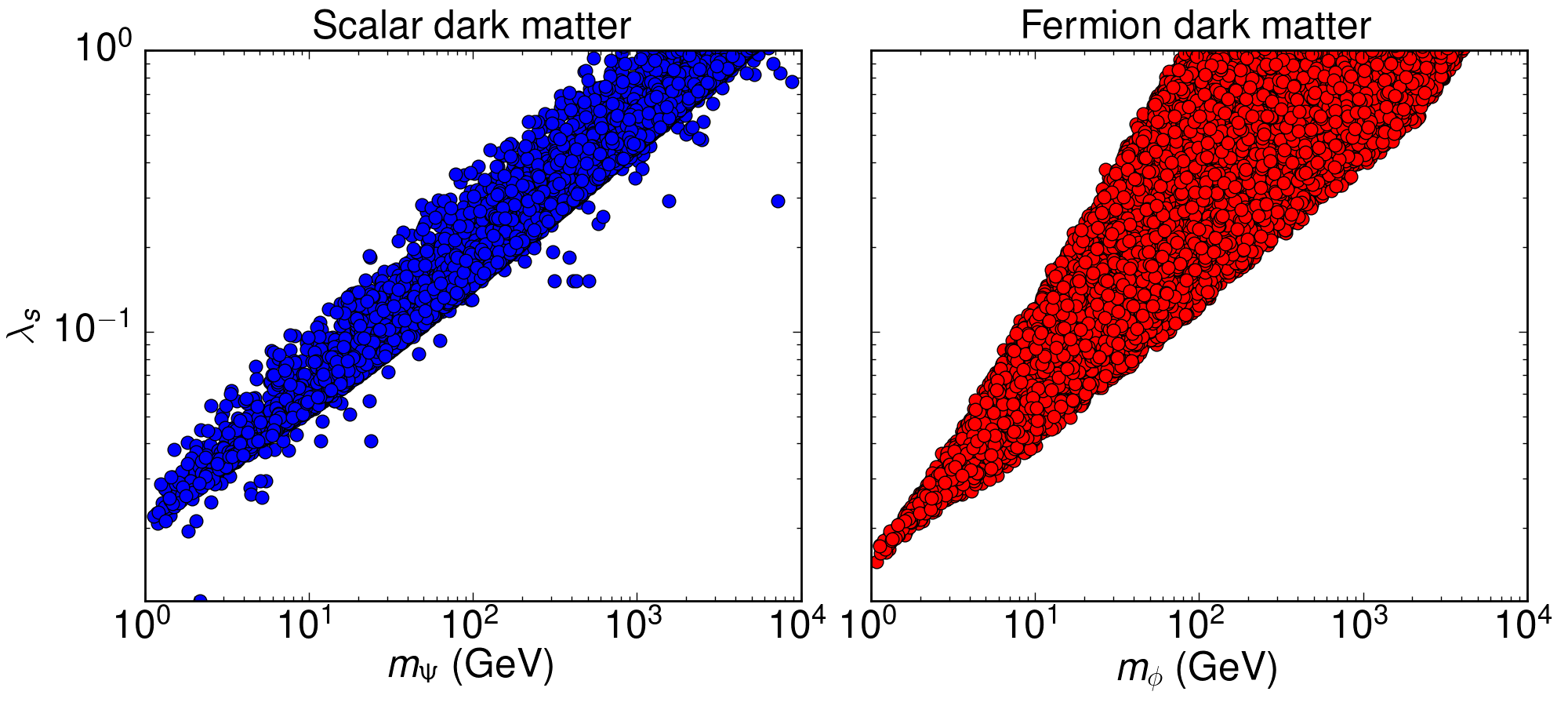}
 \caption{Allowed parameter space of the mediator mass and coupling in the scalar (left) and fermion Dark Matter (right) cases.}\label{fig:Abundance}
\end{figure}

In the following we discuss two representative cases where strong constraints can be set on the 
 parameter space of the sterile neutrino portal, namely cases where the cross section is s-wave even for  $m_N = 0$. 
We choose two benchmark scenarios, namely Majorana DM and real scalar DM with scalar couplings 
 $\alpha=\beta=\lambda_s^2$. In Fig.~\ref{fig:Abundance} we show the allowed parameter space in the mass of the mediator versus coupling, $\lambda_s$. Besides perturbativity limits, the coupling $\lambda_s$ is constrained by the width of the mediator. In our approach, the mediator particle is treated as a narrow resonance, i.e. $\Gamma/m \ll$  1, which implies $\lambda_s \lesssim \sqrt{8 \pi}$.   Taking into account this limit, these plots show that the mass of the mediator must be below $m \lesssim 1$ TeV to satisfy $\Gamma \lesssim 0.1 m$.

\begin{figure}[t!]
\begin{center}
\includegraphics[width=0.8\textwidth]{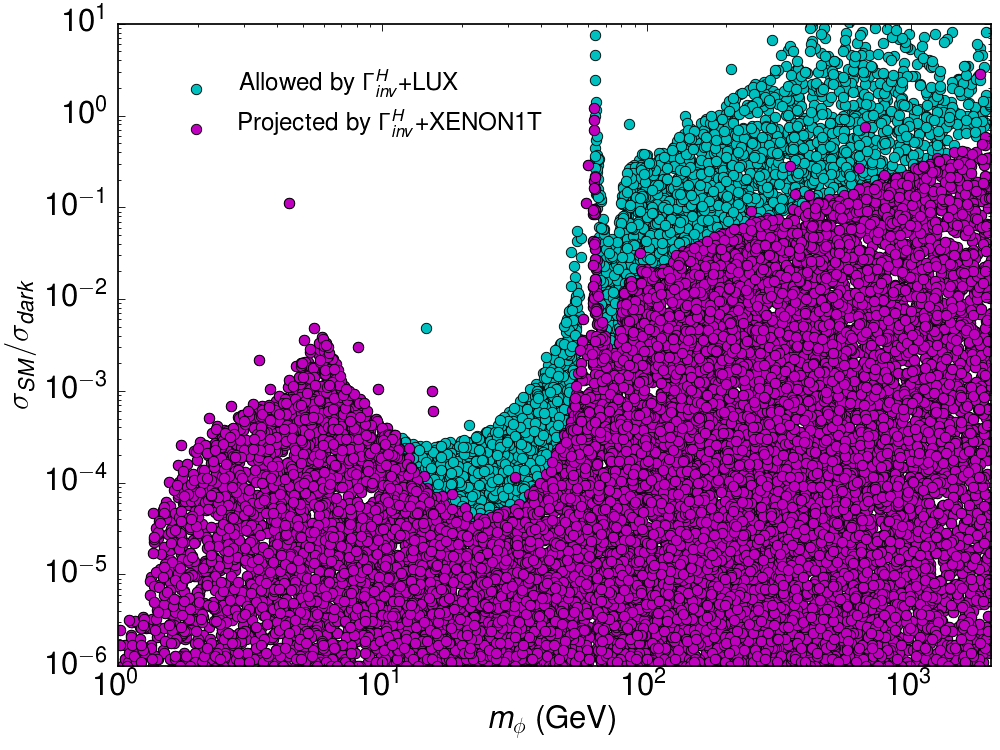}\label{fig:Brratio}
\caption{
Ratio between the cross section with Standard model particles in the final state and 
sterile neutrinos 
in the final state at $v=10^{-3}\, c$, as relevant for indirect detection searches. Currently, on the resonance $m_\phi \simeq m_h/2$ and for $m_\phi>150$ GeV 
both cross sections 
 are comparable. However,  XENON1T could set the annihilation cross section to right handed neutrinos to be dominant in the entire parameter space but for the resonance. }\label{fig:Br}
\end{center}
\end{figure}

In the scalar Dark Matter case, annihilation into right-handed neutrinos (Eq.~(\ref{eq:rel2})) is complemented via the Higgs portal coupling $\lambda_{H\phi}$ into SM particles. Namely $b \bar b$ for low mass DM, and gauge bosons and Higgses for heavier DM particles. These channels could, in principle, compete with the annihilation into right-handed neutrinos, yet in Fig.~\ref{fig:DirectDetectionS} we showed how couplings to SM are strongly constrained by direct detection experiments (LUX) and LHC bounds on the invisible width of the Higgs. We find that for $m_\phi \lesssim 100 \, {\rm GeV}$ the production cannot proceed via SM particles.
As a result of these bounds on the scalar portal, the relic abundance cannot be satisfied in the standard scalar Dark Matter, which leads to the conclusion that Higgs portal Dark Matter is not a viable scenario for low dark matter masses. This is not the case here, as our scalar has additional annihilation channels, via the coupling to dark fermions. One can then find viable scenarios, shown in Fig.~\ref{fig:Abundance}, which  satisfy the relic abundance and evade direct detection constraints in all the Dark Matter mass range from 1 GeV to 2 TeV.

Moreover, in the low dark matter mass region, annihilations to right-handed neutrinos are dominant. This is shown in  Fig.~\ref{fig:Br}, where we plot the ratio of annihilation cross sections via the Higgs portal and to the right-handed neutrino channel today, for relative Dark Matter velocity  $v=10^{-3}\, c$.
This ratio is very small, of the order or below 0.1\% for low mass, and up to 100\% for $m_\phi \gtrsim 300\, {\rm GeV}$. When the dark matter mass is low, the regions with larger  ratios are correlated with degeneracies in the dark sector, namely regions where the dark fermion mediator and the scalar are close in mass. 
This fact has implications in the ability of detecting Dark Matter today, which  
we discuss  in detail in the next section.
Notice that the $\sim 100\%$ contributions to the Dark Matter abundance currently 
allowed through SM interactions for $m_\phi \gtrsim 300 \, {\rm GeV}$, could be restricted by  XENON1T to $\sim$ 10\% for most parameter space~\footnote{The features in the low mass region of the plots are due to the fact that the contributions to the SM are mediated by the Higgs and there are several suppressions when different channels become kinematically unaccessible, $m_\phi < m_b, \, m_\tau , \, m_c$ there is a suppression factor of $(m_b/m_\tau)^2 \sim 4, \, (m_\tau/m_c)^2 \sim 2, \, (m_c/m_s)^2 \sim 100$ respectively.}. 

Finally, in the fermion Dark Matter case, since the coupling to the Higgs is generated at 1-Loop, the contributions to the annihilation cross section from the SM particles is only non-negligible in the resonant region $m_\Psi \simeq m_h/2$. 

\section{Constraints from indirect searches and CMB}\label{sec:ID}

In this scenario the annihilation of Dark Matter (with $m_{DM} \lesssim 100 \, {\rm GeV}$) into right-handed neutrinos is dominant, with the heavy neutrinos decaying into SM particles via their mixing with active neutrinos. Those decays can lead to significant fluxes of gamma rays and neutrinos which can be probed by experiments. In this section we consider the impact on the model by limits from Fermi-LAT and H.E.S.S. on the gamma-ray flux from
dwarf spheroidal galaxies~\cite{Ackermann:2015zua}
and  the galactic center~\cite{::2016jja} respectively, as well as from studies of the CMB~\cite{Slatyer:2015jla}  and IceCUBE analysis of neutrino fluxes \cite{Aartsen:2013dxa, Aartsen:2014hva, Aartsen:2015xej}.

To study the indirect detection signals in this model we first need to understand how the heavy neutrino decays. If the neutrino is light, $m_N < m_W$, $N$ will mostly decay through off-shell $Z$ and $W$. These three-body partial widths can be read from Ref.~\cite{GonzalezGarcia:1990fb,Dittmar:1989yg} and and are listed in the appendix; here we just quote the typical form it adopts:
  \bea
  \Gamma(N\to 3 \, f) \approx \frac{G_F^2}{192 \, \pi^3} \, |U_{\alpha N}|^2 \, m_{N}^5 , 
  \eea
  where $U_{\alpha N}$ is the mixing matrix between the heavy and active neutrinos. For heavier $N$, the two-body decays into massive vector bosons or Higgs and fermions are open. In this case the partial width scales as  
  \cite{Pilaftsis:1991ug}:
  \bea
  \Gamma(N\to V f) \approx \frac{g^2}{64 \, \pi m_{W,Z,h}^2} \, |U_{\alpha N}|^2 \, m_{N}^3 .
  \eea
   See also appendix A for the detailed formulae.

 The relative weight of the different lepton flavours to the total width depends on the model for neutrino mass generation. 
The large  angle $\theta_{23}$ in the active neutrino mixing matrix $U_{PMNS}$ suggests  a similar decay rate of $N$ into $\mu$ and $\tau$, while the one into $e$ is largely unconstrained.
In fact, the measured  mixing pattern (see for instance \cite{Gonzalez-Garcia:2014bfa})
 is close to Tri-Bimaximal, which leads to an exact $\mu - \tau$ symmetry \cite{Altarelli:2012ss}.
In our case, if we assume that the largest active neutrino mass  is generated by only one of the sterile neutrinos, 
$m_{3} \approx \sum_\alpha (Y_{\alpha N} v_H)^2/m_N$ and  the mixing angles are given by 
 $U_{\alpha N} \sim Y_{\alpha N} v_H/m_N$. Then, 
 $\tan \theta_{23} \sim Y_{\mu N}/Y_{\tau N} \sim 1$ and 
$\tan \theta_{13} = Y_{e N}/\sqrt{ Y_{\mu N}^2 + Y_{\tau N}^2} \sim 0.15$ 
imply that $U_{e N} \ll U_{\mu N} \approx  U_{\tau N}$ \cite{King:1999mb}.

\begin{figure}[t!]
\includegraphics[width=0.98\textwidth]{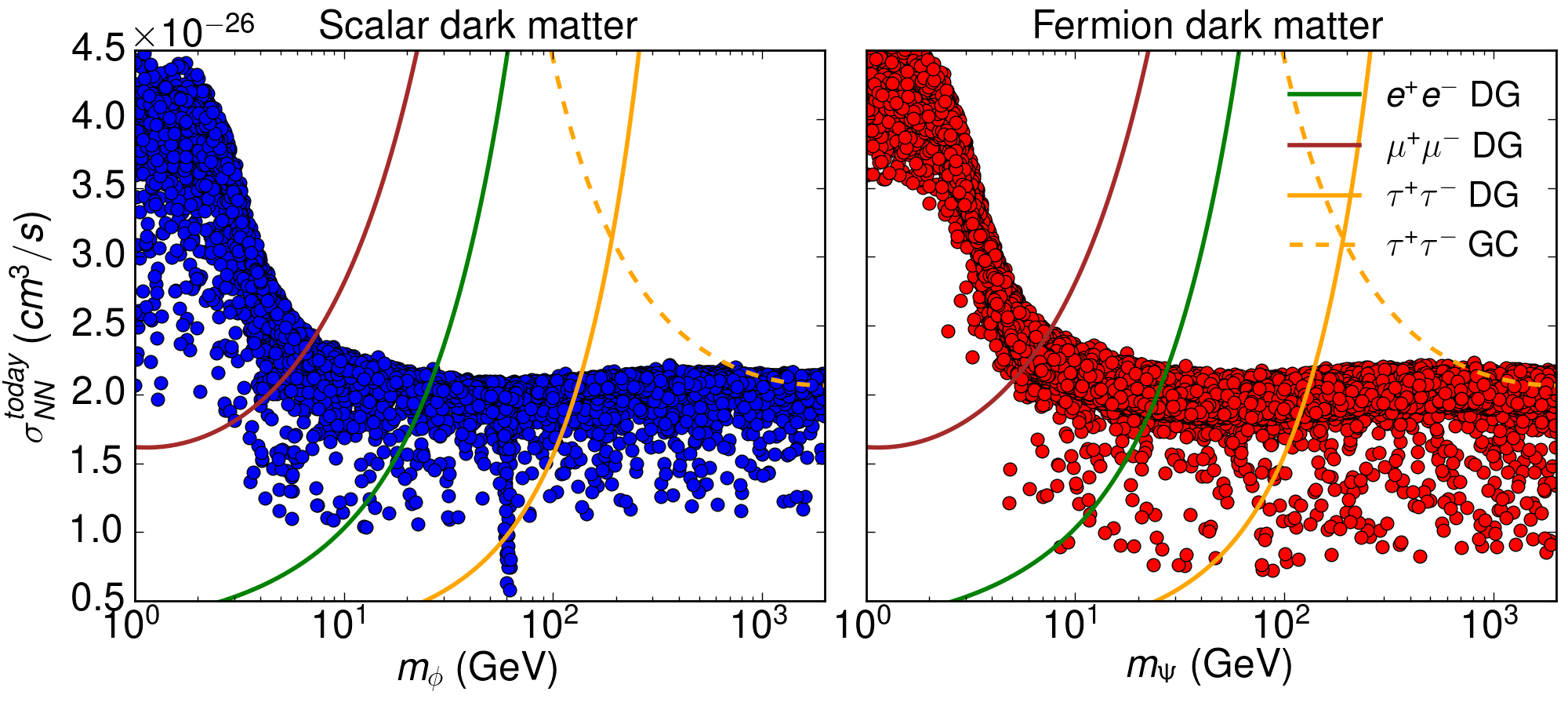} 
 \caption{Annihilation cross section today and lines of exclusion of decays to leptons from Fermi-LAT from dwarf galaxies and H.E.S.S. from the galactic center, for the cases of scalar (left) and fermionic (right) Dark Matter.}
\label{fig:IndirectFermion}
\end{figure}

A detailed study of the indirect detection signatures of our scenario is beyond the scope of this work, since DM does not decay directly to SM particles, as it is usually assumed in most analysis. Therefore we just estimate here the expected constraints using current analysis, taking into account that in general the cascade decays lead to a softer energy spectrum of the final SM particles than in the standard two body decay.
  In Fig.~\ref{fig:IndirectFermion} we present the results of such an estimate exercise in the case of decays to leptons, where limits from Refs.~\cite{Ackermann:2015zua,::2016jja} have been naively re-scaled as $m_{DM}\to m_{DM}/2$.
  We find  that decays of the right-handed neutrinos resulting into tau-leptons, e.g. from $N\to \tau q q'$ or $N\to \nu \tau^+ \tau^-$, are  potentially the most sensitive modes. Indeed, if these decays were dominant one could obtain a limit from indirect detection on the Dark Matter mass of ${\cal O}(100)$ GeV for both fermion and scalar Dark Matter. One could also use the production of quarks from off-shell $W$ and $Z$ to set bounds on the model.

Note that indirect detection signals in this case (i.e., $m_N < m_W$) have been studied in 
   \cite{Tang:2015coo}, showing that it could be possible to explain the  galactic center gamma-ray excess
   revealed by various studies of the Fermi-LAT data in 1-4 GeV gamma rays.
   Indeed, assuming that DM particles annihilate into two sterile neutrinos lighter than the $W$ boson, 
  they find that $m_N$ in the range 10 GeV to 60 GeV can explain the observed spectrum,
  while the fitted annihilation cross section $\langle \sigma v \rangle$ is (0.5 - 5) 
  $\times    10^{-26} {\rm cm^3/s}$, roughly compatible with the WIMP annihilation 
  cross section $\langle \sigma v \rangle_{\rm decouple} \sim (2 - 3 ) 
  \times    10^{-26} {\rm cm^3/s}$, when the Dark Matter particles decouple. 
More precisely, the best fit points are around $m_N \sim 30$ GeV and $m_{DM} \sim 45$ GeV,
which are within the ranges we have found compatible with all current experimental constraints
in our model. 

Finally, let us mention other sources of indirect constraints for this model.  Measurements of the cosmic microwave background (CMB) anisotropies are also sensitive to Dark Matter annihilation during the cosmic dark ages, because the injection of ionizing particles will increase the residual ionization fraction, broadening the last scattering surface and modifying  the anisotropies.
    Under the assumption that the power deposited to the gas is directly proportional to that injected at the same redshift, with some efficiency factor $f_\text{eff}$, constraints can be placed on the 
  combination $f_\text{eff} \langle \sigma v \rangle/m_{DM}$, for different SM annihilation channels in
s wave.  
  Again, the available calculations of $f_\text{eff}$ assume that DM annihilates directly to a pair
  of SM particles~\cite{Slatyer:2015jla}, and thus they are not directly applicable to our model, but we can roughly estimate 
  the expected impact of such limits in the allowed parameter space  assuming as before that  the constraints   will be similar for cascade decays, appropriately re-scaled for $m_{DM}/2$.
    Under these circumstances, we find these limits are weaker than the ones from 
   Fermi-LAT discussed above. 
      
   Besides signatures from gamma-rays, 
    in the $N N$ annihilation channel  also light neutrinos are copiously produced, which could generate an 
    observable flux from dense regions of Dark Matter. 
     IceCUBE has set constraints on the Dark Matter annihilation cross section to neutrinos by measuring the flux from 
     nearby Galaxies and Clusters~\cite{Aartsen:2013dxa}, the Galactic Halo~\cite{Aartsen:2014hva} and the Galactic Center~\cite{Aartsen:2015xej}.  However, currently these probes lie three orders of magnitude above the model prediction, and thus cannot place a constrain on our model. 
Dark Matter particles in the galactic halo can also  scatter elastically with a nucleus and become trapped in the gravitational well of astronomical objects like the Sun, eventually thermalize and concentrate at the core of the object. Then, they may annihilate into SM particles, in particular neutrinos that can be detected by neutrino experiments such as IceCUBE or SuperKamiokande. 
  In our scenario the limits from direct searches are tighter than such indirect probes, since the interaction of Dark Matter to quarks is spin independent.

\section{Conclusions and outlook}~\label{sec:Concls}
In this paper we have analyzed in detail a simple scenario of a dark sector composed of a scalar and a fermion, both singlets under the SM 
gauge group but charged under a dark symmetry group. This sector is linked to the origin of neutrino masses via couplings to the sterile neutrinos, which are able to mediate between the dark sector and the SM. 

This scenario has been studied in Refs.\cite{Macias:2015cna,Gonzalez-Macias:2016vxy}, considering just the case of fermionic Dark Matter and for sterile neutrinos heavier than the dark sector, with the result that 
current experimental and observational constraints 
(electroweak precision limits, Dark Matter relic abundance,  direct and indirect detection constraints), 
can be accommodated only for  $m_{DM} \lesssim$ 35 GeV, or in the resonances, $m_{DM} \simeq m_h,m_Z$,
unless the dark scalar or dark fermion are almost degenerate. 

We have extended these previous studies in two ways: we explore the phenomenology 
of this type of models 
when the sterile neutrinos are lighter than the dark sector, so that the Dark Matter annihilation channel into $N N$ is kinematically allowed, 
 and we consider both,  fermionic and scalar Dark Matter in this context.
We have 
performed for the first time an exhaustive numerical analysis of this alternative region of the model parameter space, 
and after imposing all the relevant constraints  from direct detection and  collider probes, 
we find that 
it is possible to obtain the observed Dark Matter relic abundance in the whole mass range explored, 
$m_{DM} \in \, [{\rm 1 \, GeV, 2 \, TeV}]$, both for scalar and fermion Dark Matter. 

We find that the scalar case is an interesting extension of the Higgs portal. Indeed, in the usual portal the constraints on the Higgs invisible decay and Dark Matter nucleon cross section rule out the possibility of the scalar as the main component of Dark Matter for $m_\phi \lesssim 100 \, {\rm GeV}$. But in our scenario, annihilation can occur via the neutrino portal which is dominant, i.e. more than 90\%, in most of the parameter space. On the other hand, in the case of a fermion Dark Matter, the contribution to the quark-Dark Matter scattering and Higgs invisible width decay is at one-loop and the Higgs portal coupling is only mildly constrained. 

Finally we explore the indirect detection characteristics of this model, determined by the decays of the right-handed neutrinos into SM bosons and leptons. We consider constraints from Fermi-LAT and find that those could be sensitive to Dark Matter up to the electroweak scale, $m_{\rm DM} \lesssim 100 \, {\rm GeV}$ independently of wether the Dark Matter particle is a scalar of a fermion. However, a more detailed analysis of these constraints need to be done, as we performed a naive scaling on constraints of Dark Matter decays to two SM particles. In our scenario, the more complex decays of right-handed neutrinos would lead to less energetic SM probes. Finally, we also comment on the possibility of this channel to be responsible of the gamma ray galactic excess at few GeV.

 \section*{Acknowledgements} 
We thank Olga Mena, Sergio Palomares Ruiz, Roberto Ruiz de Austri, 
Jordi Salvad\'o, A. Vincent and J. Wudka for illuminating discussions, and Concha Gonz\'alez-Garc\'\i a 
for comments on the manuscript.
ME thanks Antonia Abenza for inspiring and encouraging conversations.
 This work has been partially supported by the European UnionÕs Horizon 2020 research and innovation programme under the Marie Sklodowska-Curie grant agreements No 674896 and 690575, 
by the Spanish MINECO under grants FPA2014-57816-P and  SEV-2014-0398,
and by Generalitat Valenciana grant PROMETEO/2014/050.
ME is supported by Spanish Grant FPU13/03111 of MECD. 
NR acknowledges the support from the Munich Institute for Astro- and Particle Physics (MIAPP) of the DFG cluster of excellence ``Origin and Structure of the Universe".
The work of VS is supported by the Science Technology and Facilities Council (STFC) under grant number ST/J000477/1.

\appendix

\section{Sterile neutrino decay widths}

Here we summarize the sterile neutrino decay modes, relevant for indirect Dark Matter searches.

  If the sterile neutrino  is lighter than the $W$ boson, it will decay through off-shell $h,Z,W$
 bosons to three fermions. Since the decay via a virtual $h$ is further suppressed by the small Yukawa couplings of the SM fermions, it is a very good approximation to consider only the processes mediated by virtual $W,Z$, whose partial widths read~\cite{GonzalezGarcia:1990fb}:
\bea 
 \Gamma(N  \rightarrow \nu q \bar{q} ) &=& 3\, A C_{NN}
[2(a_u^2 + b_u^2) + 3(a_d^2 + b_d^2)] f(z) 
\\
 \Gamma(N   \rightarrow  3 \nu) &=& A C_{NN}
[\frac 3 4 f(z)  +  \frac 1 4  g(z,z)]  
\\
 \Gamma(N   \rightarrow \ell q \bar{q} ) &=& 6\, A C_{NN}
 f(w,0) 
\\
 \Gamma(N   \rightarrow \nu \ell\bar{\ell} ) &=&  A C_{NN}
[3(a_e^2 + b_e^2) f(z)  + 3 f(w) - 2 a_e g(z,w) ]
\eea
where $C_{NN}$ is defined in Eq.~(\ref{eq:cij}), 
\be
A \equiv  \frac {G_F^2 m_{N}^5 }{192 \, \pi^3}  \ ,
\ee
$a_f,b_f$ are the left and right neutral current couplings of the fermions ($f=q,\ell$),  
the variables $z,w$ are given by 
\be 
z= (m_N/m_Z)^2  \ , \qquad w = (m_N/m_W)^2 \ , 
\ee 
 and the functions $f(z), f(w,0)$ and $g(z,w)$ can be found in \cite{Dittmar:1989yg}.

 For larger values of $m_{N}$, two body decays to SM particles are open, and the corresponding widths 
 read \cite{Pilaftsis:1991ug}:
  \bea 
  \label{eq:WZh}
  \Gamma(N  \rightarrow W^{\pm} \ell^{\mp}_{\alpha}  ) &=& \frac{g^2}{64 \pi} |U_{\alpha N}|^2 
  \frac{m_N^3}{m_W^2} \left(1 -   \frac{m_W^2}{m_N^2} \right)^2
   \left(1 +  \frac{2 m_W^2}{m_N^2} \right)  
   \\
    \Gamma(N  \rightarrow Z  \, \nu_\alpha ) &=& \frac{g^2}{64 \pi c_W^2 } |C_{\alpha N}|^2 
  \frac{m_N^3}{m_Z^2} \left(1 -   \frac{m_Z^2}{m_N^2} \right)^2
   \left(1 +  \frac{2 m_Z^2}{m_N^2} \right)    
   \\
   \Gamma(N  \rightarrow h \,  \nu_\alpha ) &=& \frac{g^2}{64 \pi } |C_{\alpha N}|^2 
  \frac{m_N^3}{m_W^2} \left(1 -   \frac{m_h^2}{m_N^2} \right)^2  
  \eea 
 
  In the above expressions, we have assumed that $N$ is a Majorana fermion. If it is Dirac, then the 
 decay channel $N  \rightarrow W^{-} \ell^{+}  $ is forbidden and the decay widths
 into $Z/h, \nu$ are 
 $ \Gamma_D (N  \rightarrow Z/h \ \nu_\ell ) =   \Gamma_M (N  \rightarrow Z/h  \ \nu_\ell )/2$.

 \bibliography{DM}
 \bibliographystyle{JHEP}

 \end{document}